\documentclass[nofootinbib,prd,12pt,superscriptaddress]{revtex4}%
\usepackage{amsmath}
\usepackage{amsfonts}
\usepackage{amssymb}
\usepackage{graphicx}%
\usepackage{amsmath,amssymb}
\usepackage{mathrsfs}
\usepackage{graphicx}
\usepackage{color}
\usepackage{subfigure}
\usepackage{fancyhdr}
\usepackage{multirow}
\usepackage{float}
\usepackage{epsfig}
\usepackage{amsfonts}
\usepackage{bm}

\def\be{\begin{equation}}
\def\ee{\end{equation}}
\def\beq{\begin{eqnarray}}
\def\eeq{\end{eqnarray}}

\begin{document}

\title{Exploring Self-Gravitating Cylindrical Structures in Modified Gravity: Insights from Scalar-Vector-Tensor Theory}
\author{Davood  Momeni}
\affiliation{Northeast Community College, 801 E Benjamin Ave Norfolk, NE 68701, USA}

\author{Phongpichit Channuie}
\affiliation{College of Graduate Studies, Walailak University, Nakhon Si Thammarat, 80160, Thailand}
\affiliation{School of Science, Walailak University, Nakhon Si Thammarat, 80160, Thailand} 

\author{Mudhahir Al-Ajmi}
\affiliation{Physics Department, College of Science, P.O. Box 36, Sultan Qaboos University 123, Oman}
\begin{abstract}

We investigate static cylindrical solutions within an extended theory of modified gravity. By incorporating various coupling functions through a straightforward boost symmetry approach, we establish the equations of motion in a self-consistent manner and subsequently determine the linear scalar field profile. Utilizing analytical methods, we solve the system of equations for the metric functions and the $U(1)$ gauge field, revealing their dependence on Bessel's functions. To comprehend gravito-objects exhibiting cylindrical symmetry, we develop a perturbative framework aimed at identifying all nontrivial solutions for the scalar profiles. Introducing first-order truncated perturbation equations for the gauge field, synchronized with metric gauges and electromagnetic field considerations, we demonstrate their integrability and obtain solutions through quadrature. Our findings suggest the feasibility of obtaining self-gravitating cylindrical structures within the scalar-vector-tensor theory. These cylindrical structures could provide insights into the behavior of gravitational and gauge fields in modified gravity, potentially offering new perspectives on astrophysical phenomena such as cosmic strings and cylindrical gravitational waves.
\end{abstract}


\maketitle

\section{Introduction}
In physics, fundamental forces are often formulated using gauge theory, allowing us to describe interactions between matter and fields through perturbation theory. This approach, known as the $S$-matrices formalism, enables computation of scattering amplitudes and cross sections, providing insights into the structure of matter. In the realm of four-dimensional spacetime, general relativity stands as a fully general covariant theory, sharing commonalities with classical gauge theories such as electromagnetism.

While Einstein's theory of gravity remains highly successful in explaining gravitational phenomena, the need to extend general relativity arises to address various challenges, including the cosmological constant problem, dark energy, dark matter, and the quest for a quantization scheme. One straightforward extension involves introducing a scalar field, known as the scaleron degree of freedom. However, such modifications may introduce higher-order time derivatives in the equations of motion, potentially leading to Ostrogradski instabilities \cite{Heisenberg:2018acv}.

In certain cases, these higher-order time derivatives can be formally transformed into standard second-order forms, as demonstrated by Horndeski \cite{Horndeski:1974wa}, \cite{Horndeski:1976gi}. He showed that the most general second-order, covariant scalar-tensor theories lead to stable Horndeski interactions with specific coupling functions \cite{Horndeski:1976gi}. Additionally, the inclusion of an auxiliary vector field in modified theories, meaningfully coupled to other aspects of the gravitational sector, leads to the most general vector-tensor theories with second-order equations of motion. These extensions, following Horndeski's terminology, result in specific viable couplings between the vector field and the curvature tensor, reminiscent of Maxwell's equations under carefully chosen gauge fixing schemes. Relaxing these gauge fixings leads to a broader class of theories known as generalized Proca theories.

Further developments have identified novel purely intrinsic vector interactions without scalar counterparts, such as non-minimally coupled theories to the double dual Riemann tensor, with significant cosmological and astrophysical implications. The scalar-vector-tensor (SVT) theory emerges as a unification of Horndeski scalar-tensor theories and generalized Proca theories, motivated by recent works \cite{Heisenberg:2014rta}. This class of modified gravity theories presents new solutions for hairy black holes and offers alternative scenarios for dark energy \cite{Jimenez:2016isa}-\cite{Tsujikawa:2015upa}.

Previous studies have primarily focused on spherical solutions in modified gravity theories, which have provided significant insights but also revealed certain limitations. Spherical solutions are often idealized and may not capture the full complexity of various astrophysical and cosmological structures. For instance, while spherical symmetry simplifies the mathematical treatment, it may not adequately describe elongated or cylindrical objects observed in the universe, such as cosmic strings or certain types of gravitational waves. Addressing these limitations necessitates exploring solutions with different symmetries.

In this context, cylindrical solutions offer a compelling avenue for further research. Cylindrical structures can model various astrophysical phenomena and provide a richer framework for understanding gravitational interactions in modified gravity theories. Additionally, previous studies on cylindrical solutions in modified gravity have been undertaken. For a more thorough background, we direct readers to Refs. \cite{Houndjo:2012sz, Houndjo:2013us, Momeni:2015aea, Azadi:2008qu}. These studies underscore the importance of extending our investigations to cylindrical symmetry to fully grasp the potential implications of modified gravity theories.

In this paper, we focus on investigating static, non-rotating cylindrical solutions within the SVT framework. The paper is organized as follows: Section 2 provides a brief overview of the modified scalar-vector-tensor theories of gravity. In Section 3, we propose a static Weyl's gauge suitable for describing cylindrical objects. Section 4 is dedicated to deriving equations of motion for all fields, while Section 5 presents exact solutions for linear and quadratic scalaron profiles. Section 6 discusses the integrability of perturbative equations of motion in detail, and the final section offers concluding remarks.
\section{Heisenberg's Gauge-invariant scalar-vector-tensor theories }\label{s2}
SVT theory offers a comprehensive framework to address various cosmological enigmas such as dark energy, dark matter, and the accelerated expansion of the universe. One of the primary advantages of SVT theory over other modified gravity theories is its ability to naturally incorporate both scalar and vector degrees of freedom, leading to richer and more versatile interactions. This versatility allows the SVT theory to simultaneously account for the large-scale structure of the universe and local gravitational phenomena without introducing instabilities. Additionally, SVT theory can produce viable cosmological models that align with observational data, offering alternative explanations for the phenomena typically attributed to dark energy and dark matter. By unifying aspects of Horndeski scalar-tensor theories and generalized Proca theories, SVT theory provides a robust platform for exploring new gravitational dynamics and potentially resolving long-standing cosmological puzzles.
\par
The objective of this section is to provide a foundational framework for understanding the Lagrangian of the innovative modified gravity theories recently proposed in Ref. [1]. Referred to as scalar-vector-tensor gravity theories, this class of models is widely regarded as a fusion of Horndeski theories and generalized Proca theories. The SVT framework exhibits gauge dependency and necessitates meticulous gauge fixing. By adopting a suitable gauge-invariant platform, one arrives at the model introduced in Ref.\cite{Heisenberg:2018acv}.
The Lagrangian densities of typical scalar-vector-tensor theories, incorporating arbitrary interaction terms (coupling functions), can be represented as follows:
\begin{eqnarray}\label{genLagrangianSVTnoGauge}
\mathcal{L}^{2}_{\rm SVT}&=&f_2(\pi,X,F,\tilde{F},Y), \\
\mathcal{L}^{3}_{\rm SVT}&=& {\cal M}_{3}^{\mu\nu}\nabla_\mu\nabla_\nu\pi,\label{eq2}\\
\mathcal L^{4}_{\rm SVT} & = & {\cal M}_{4}^{\mu\nu\alpha\beta}\nabla_\mu\nabla_\alpha\pi\nabla_\nu\nabla_\beta\pi+f_{4}(\pi,X)L^{\mu\nu\alpha\beta}F_{\mu\nu}F_{\alpha\beta},\label{eq3}
\end{eqnarray}
In the above Lagrangian, the  functions $X=-\frac12(\partial\pi)^2$ (kinetic term), $F=-F_{\mu\nu}F^{\mu\nu}/4$  , ${\tilde F}=-F_{\mu\nu}{\tilde F}^{\mu\nu}/4$   and $Y=\nabla_\mu\pi\nabla_\nu\pi F^{\mu\alpha}F^\nu{}_\alpha$ (field strength terms) with the gauge-invariant field strength and its dual given by $F_{\mu\nu}=\nabla_{\mu}A_{\nu}-\nabla_{\nu}A_{\mu}$ and ${\tilde F}_{\mu\nu}=\epsilon^{\mu\nu\alpha\beta}F_{\alpha\beta}/2$, respectively. In the standard notations,  $\epsilon^{\mu\nu\alpha\beta}$ is the anti-symmetric Levi-Civita tensor. The auxiliary rank-2 tensor
${\cal M}^{\mu\nu}_{3}$ in Eq.(\ref{eq2}) is defined as 
\begin{eqnarray}
{\cal M}^{\mu\nu}_{3}= \Big(f_{3}(\pi,X)g_{\rho\sigma}+{\bar f}_{3}(\pi,X)\nabla_{\rho}\pi\nabla_{\sigma}\pi\Big)\tilde{F}^{\mu\rho}\tilde{F}^{\nu\sigma}\,,
\end{eqnarray}
where $f_{3}$ and ${\bar f}_{3}$ are functions of $\pi$ and $X$. Furthermore another auxiliary  rank-4 tensor  $\mathcal{M}_{4}^{\mu\nu\alpha\beta}$ in Eq.(\ref{eq3}) is expressed  as
\begin{equation}
\mathcal{M}_{4}^{\mu\nu\alpha\beta}=\Big( \frac12f_{4,X}(\pi,X)+{\tilde f}_{4}(\pi)\Big)\tilde{F}^{\mu\nu}\tilde{F}^{\alpha\beta}\,.
\end{equation}
In the above relation, the  $L^{\mu\nu\alpha\beta}$ is the double dual Riemann tensor and is formulated  with the help of the Riemann tensor $R_{\rho\sigma\gamma\delta}$ as
\begin{eqnarray}
L^{\mu\nu\alpha\beta}=\frac{1}{4} \epsilon^{\mu\nu\rho\sigma}\epsilon^{\alpha\beta\gamma\delta} R_{\rho\sigma\gamma\delta}\,,
\end{eqnarray}

The renowned SVT framework offers a diverse landscape for formulating and exploring theories that extend beyond scalar-vector-tensor formulations. This can be achieved by implementing disformal transformations \cite{Tsujikawa:2015upa}. A plausible and physically meaningful interpretation of the interactions in ${\cal L}^{2,3,4}_{\rm SVT}$ is to view them as generators of disformal transformations or coupling functions \cite{disformal}.

In the subsequent sections, we will delve into how the current theories, up to ${\cal L}^{3}_{\rm SVT}$, yield specific geometries conducive to cylindrical objects.

\section{Cylindrical geometry}
\label{s3}
We initiate our investigation by formulating the general cylindrical spacetime metric in Weyl's coordinates, expressed as follows:
\begin{equation}\label{metric}
ds^2=g_{\mu \nu}dx^\mu dx^\nu = e^{2U} dt^2 -e^{-2U}[e^{2K}(d\rho^2+dz^2)+r^2 d\varphi]\,,    \end{equation}
Here, $x^{\mu}=(t,r,\varphi,z)$ represent Weyl's cylindrical coordinates. The spacetime exhibits trivial Killing symmetry generators $\zeta^{t}=\partial_t$ and $\zeta^{\varphi}=\partial_{\varphi}$. We focus on time-independent (stationary) non-rotating (static) mass distributions, leading to a diagonal metric $g_{\mu\nu}$. Notably, the metric provided in eq. (\ref{metric}) resembles that of a cosmic string. Despite residing in flat spacetime, the azimuthal angle $\varphi\not\in(0,2\pi]$, yet the local geometry remains akin to flat spacetime.

Adhering to this symmetry class, we assume all metric functions to be solely functions of $r$, denoted as $U=U(r)$ and $K=K(r)$. Exact vacuum solutions to the Einstein field equations with this symmetry can be attained, leading to the discovery of the Levi-Civita solution. Extensive investigations into cylindrical solutions across different gravitational theories have been conducted in the past, primarily aimed at understanding topological defects through Riemannian geometry \cite{vilenkin}. Inspired by string theory, this approach elucidates how topological defects mimic cylindrical solutions.

In general relativity (GR), the simplest cylindrical model, presented as an exact class of metrics, was discovered by Kasner in Refs. \cite{Kasner}. Linet subsequently proposed the first extension of the Kasner solution in an exact form \cite{Linet}. Tian later generalized the Linet solution by incorporating a cosmological constant \cite{tian}, offering the capability to study both cosmological and AdS limits in cylindrical systems. Within the framework of modified gravity theories, the first extension of the GR solution with cylindrical symmetry in $f(R)$ gravity was presented in \cite{fr}. Matter contents can be weakly coupled to the gravitational sector without violating the equivalence principle, as studied as a source for cosmic strings by Harko in Ref. \cite{frlm}.

Given the origin of these solutions in string theory, higher-order corrections to the GR solutions have garnered importance, such as the Gauss-Bonnet corrections to cosmic strings investigated in \cite{gb}. When Lorentz symmetry is broken in high-energy regimes, non-relativistic theories emerge as alternatives to GR, modifying cylindrical geometries \cite{nonrelativistic}. Harko and Lake made a remarkable observation, discovering a naive relation between the geometries of cosmic strings and the Bose-Einstein condensed phase of matter \cite{be}. Further corrections to GR, such as mimetic complements to GR, also support the existence of cylindrical gravitational objects \cite{mimetic}. In teleparallel theories, where gravity is viewed as the effect of torsion rather than curvature, it is also feasible to find metrics for cosmic strings \cite{teleparallel}. Moreover, cylindrical solutions have been studied in various contexts, including brane-world scenarios \cite{brane}, Kaluza-Klein models \cite{kk}, Lovelock Lagrangians \cite{love}, Born-Infeld \cite{bi}, bimetric theories \cite{bit}, scalar-tensor theories \cite{st}, Brans-Dicke theory \cite{bd}, and dilation gravity \cite{dilaton}.\par

As indicated in Ref. \cite{Heisenberg:2018acv}, the SVT theories offer the possibility of accommodating new black object solutions, marking one of the pivotal applications of these theories. To commence our inquiry, we will first delve into the cylindrical solutions within the framework of these theories. To explore these potential solutions, let us examine the following Lagrangian, with a metric signature of $(+, -, -, -)$:
\begin{eqnarray}
&&\mathcal{L}=\frac{M_{Pl}^2}{2}R+\mathcal{L}^3_{SVT}\,,\label{Lagrangina}
\end{eqnarray}
In the given Lagrangian, $M_{\text{Pl}}^2 = 8\pi G$ represents Planck's mass, and $R$ denotes the Ricci scalar for the specified geometry.
\par
For the metric provided in Eq. (\ref{metric}), we compute the Ricci scalar curvature and the determinant of the metric as follows: 
\begin{eqnarray}
R&=&-\frac{2}{r}e^{2U-2K}(rU'^2+r(K''-U'')-V')\,,
\\
\sqrt{-g}&=& re^{2K-2U}.
\end{eqnarray}

Throughout this study, the primes represent derivatives with respect to $r$, the radial cylindrical coordinate. To maintain both stationarity and time independence, we choose the scalar field profile as:
\begin{equation}
\pi=\pi(r)\,.
\end{equation}
The corresponding kinetic term is expressed as:
\begin{equation}
X=-\frac{1}{2}e^{-2U}\pi'^2\,,
\end{equation}
We assume that the $U(1)$ gauge field has the following covariant components, taking into account its symmetry:
\begin{equation}
A_\mu=(A(r),0,0,0)\,. 
\end{equation}
Utilizing the aforementioned proposition, we find:
\begin{equation}
F=-2A'^2 e^{2K}\,.    
\end{equation}
To determine the remaining coupling functions in the SVT Lagrangian, it's important to note that the coupling function $G_2$ primarily depends on the metric functions and the gradient of the $U(1)$ field.
\begin{equation}
G_2=G_2\Big(\pi(r), -\frac{1}{2}e^{-2U}\pi'^2, -2A'^2 e^{-2K}\Big).  \label{G2}
\end{equation}
Using the aforementioned functional dependency, we can express the third-order term in the gravity sector of the SVT theory as follows:
\begin{equation}
\mathcal{L}^3_{SVT}=-\Big(f_3(\pi,x)g_{rr}+\tilde{f_3}(\pi,x)\pi'^2\Big)\pi' e^{4U-6K} U'A'^2\,. \label{L3}
\end{equation}
Finally, by substituting (\ref{L3}) into the full Lagrangian, i.e., Eq. (\ref{Lagrangina}), we obtain:
\begin{eqnarray}
\mathcal{L}&=&-M_{Pl}^2(rU'^2+r(K'-U')-U')+G_2\big(\pi,-\frac{1}{2}e^{-2U}\pi'^2,-2A'^2 e^{-2K}\big) \nonumber \\
&&-\pi'U'A' e^{4U-6U}\big(-f_3(\pi,X)e^{2K-U}+\pi'^2\tilde{f_3}(\pi,X)\big)\,.\label{Lfull}
\end{eqnarray}
We emphasize that the gravity model described by the Lagrangian in Eq. (\ref{Lfull}) remains intricate. Consequently, we must proceed with further simplifications, primarily inspired by Horndeski theories and other scalar-tensor theories. We propose the following simplified model for the coupling functions in our theory:
\begin{eqnarray}
G_2(\pi,-\frac{1}{2}e^{-2U}\pi'^2, -2A'^2e^{-2K})&=&V_1(\pi)-\frac{\alpha}{2}e^{-2U}\pi'^2-2\beta A'^2 e^{-2K}\,,  
\\
f_3(\pi,-\frac{1}{2}e^{-2U}\pi'^2)&=&V_2(\pi)-\frac{\gamma}{2}e^{-2U}\pi'^2 \,,  
\\
\tilde{f_3}(\pi,-\frac{1}{2}e^{-2U}\pi'^2)&=&V_3(\pi)-\frac{\delta}{2}e^{-2U}\pi'^2 .  
\end{eqnarray}
In the above expressions, the auxiliary potential functions, denoted as $V_i(\pi)$ for $i=1,2,3$, are ordinary potential functions. These will be determined later in the course of this work.

\section{Equations of motion}\label{s4}

In the preceding section, we derived the reduced point-like Lagrangian for the cylindrical gravitational objects, as presented in Eq. (\ref{Lfull}). To ensure clarity and rigor in deriving the equations of motion, we follow a systematic approach that integrates the boost symmetry principles. This approach not only simplifies the form of our metric and gauge field components but also underscores their physical relevance in the context of cylindrical gravitational structures. By enforcing boost symmetry, we constrain the metric functions $U$, $K$, and $\pi$ to be functions solely of the radial coordinate $r$, while allowing the gauge field $A$ to retain dependence on both $r$ and the angular coordinate $\varphi$. This reduction in degrees of freedom enhances the tractability of our equations, facilitating a more insightful exploration of their physical implications within the framework of SVT theory.
This reduced Lagrangian comprises four unknown functions denoted as $q^a={U,K,\pi,A}$. In the language of dynamics, obtaining the equations of motion (EoMs) is straightforward using the standard Euler-Lagrange equations, which are expressed as follows:
\begin{eqnarray}
\frac{d}{dr}\Big(\frac{\partial \mathcal{L}}{\partial \phi_i'}\Big)=\frac{\partial \mathcal{L}}{\partial \phi_i} ,\ \ \phi_i=(U,K,\pi,A).
\end{eqnarray}
In the set of equations above, the radial coordinate $r$ assumes the role of time in the dynamical system. It is widely acknowledged that this radial coordinate $r$ should be redefined appropriately as another physical radius coordinate to measure the distances between causal events in the spacetime manifold. An intriguing polymerization of the equations of motion is presented as follows:
\begin{eqnarray}\label{eom-U1}
U+2-4( U'+rU'') = \sum_{n=1}^{5}h_n(U,K,A',A'')(\pi'')^n,
\end{eqnarray}
In the polymerization of the equations of motion, the coefficients $h_i$ are functions of the variables $q^a$, and successive derivatives of them are defined as follows:
   \begin{eqnarray}
&&h_1(U,K,A',A'')= 4\,{{\rm e}^{2\,U  -4\,K  }}V_{{2}
} \left( \pi  \right) {\it A'}  {\it K'}   -{
{\rm e}^{2\,U  -4\,K  }}{\it A''}
  V_{{2}} \left( \pi  \right)-
  {\it A'}  {{\rm e}^{2\,U  -4\,K  }}V_{{2}} \left( \pi  \right)\,,
   \\
&&h_2(U,K,A',A'')=
\alpha\,{{\rm e}^{-2\,U  }}-{{\rm e}^{2\,U  -4\,K  }
}{\it V'_2} \left( \pi  \right) {\it A'}\,,  \\
&&h_3(U,K,A',A'')=
\frac{1}{2}
{{\rm e}^{-4\,K  }}{\it A''}
  \gamma+\frac{3}{2}\, {{\rm e}^{-4\,K  }}
\gamma\,{\it A'}-2\,{{\rm e}^{-4\,K  }}\gamma\,{\it A'}  {
\it K'}    +3\,
 {\it A'}  {{\rm e}^{4\,U  -6\,K  }}V_{{
3}} \left( \pi  \right)\nonumber\\&&
-6\,{{\rm e}^{4\,U  -6\,K
  }}V_{{3}} \left( \pi  \right) {\it A'}  {\it K'}  + {{\rm e}^{4\,U  -6\,K  }}{\it A''}  V_{{3}} \left( \pi  \right)\,,\\
  &&
h_4(U,K,A',A'')={
{\rm e}^{4\,U  -6\,K  }}{\it V'_3}
 \left( \pi  \right) {\it A'}  \,,
  \\
&&h_5(U,K,A',A'')=
  -\frac{5}{2}\, {\it A'}  {{\rm e}^{2\,U  -6\,K  }}\delta   +3\,{{\rm e}^{2\,U  
-6\,K  }}\delta\,{\it A'}  {\it K'}
     -\frac{1}{2}\, {{\rm e}^{2\,U  -6\,K
  }}{\it A''}  \delta. 
\end{eqnarray}
For another metric function $K$, we similarly have:
\begin{eqnarray}
\label{eom-K}
K-4\beta\, \left(A'\right)^{2}{{\rm e}^{-2\,K}}-2=\sum_{n=1,3,5}p_n(U,K,U',A')(\pi'')^n\,,
\end{eqnarray}
where the set of the auxiliary potential functions are defined by
\begin{eqnarray}
&&p_1(U,K,U',A')=-4U'A'e^{2U-4K}V_2(\pi)\,,
\\&&
p_3(U,K,U',A')=2\gamma U'A'e^{-4K}+6U'A'V_{3}(\pi)e^{4U-6K}V_3(\pi)\,,
\\&&
p_5(U,K,U',A')=-3\delta  U'A'e^{2U-6K}\,.
\end{eqnarray}
For the scaleron field $\pi$, the Euler-Lagrange equation takes the form of a generalized Klein-Gordon-like equation, which can be explicitly written as:
\begin{eqnarray}
\label{eom-PI}
\pi- V'_{1}(\pi)&+& A' e^{2\,U
  -4\,K  }V_{2} (\pi)\Big( U'+U''-4U'K'-2U'^2\Big)
   \nonumber\\&=& \sum_{n=1}^{4}l_{n}(U,K,A',K',U',A'',U'')(\pi'')^n\,,
  \end{eqnarray}
where the coefficients are expressed as follows:
  \begin{eqnarray}&&
  l_{1}(U,K,A',K',U',A'',U'')={{\rm e}^{-2\,U  }}\alpha-2\,{{\rm e}^{-2\,U  }}\alpha\,{\it U'}  \,,
  \\&&
  l_{2}(U,K,A',K',U',A'',U'')=3\,{{\rm e}^{-4\,K  }}
\gamma\,{\it A'}  {\it U'} -6\,{{\rm e}^{-4\,K \left( r
 \right) }}\gamma\,{\it A'}  {\it K'}  {\it U'}  \\&&\nonumber-18\,{{\rm e}^{4\,U   -6\,K \left( r
 \right) }}V_{{3}} \left( \pi  \right) {\it A'}  {\it
K'}  {\it U'}   +6\,{{\rm e}^{4\,U  -6
\,K  }}V_{{3}} \left( \pi  \right) {\it A'}  {\it U'}   +3\,{\it U''}  {\it A'} {
{\rm e}^{4\,U  -6\,K  }} V_{{3}} \left(\pi\right)
\\&&\nonumber+12\,{{\rm e}^{4\,U  -6\,K  }}V_{{3}} \left( \pi  \right) {\it A'}  
{\it U'}^{2}  +3\,{\it A''}  {\it U'}
  {{\rm e}^{4\,U  -6\,K  }} V_{{3}}
 \left( \pi  \right) +\frac{3}{2}\,{\it A''}
  {{\rm e}^{-4\,K  }}\gamma\,{\it U'}+
  \,{\it U''}  {{\rm e}^{-4\,K  }}\gamma
\,{\it A'} \,,
  \\&&
  l_{3}(U,K,A',K',U',A'',U'')=2\,{
{\rm e}^{4\,U  -6\,K  }}{\it V'_3}
 \left( \pi  \right) {\it A'}  {\it U'}  \,,
  \\&&
  l_{4}(U,K,A',K',U',A'',U'')=15\,{{\rm e}^{2\,U  -6\,K  }
}\delta\,{\it A'}  {\it K'}  {\it U'}-10
\,{{\rm e}^{2\,U  -6\,K  }}\delta\,{
\it A'}  {\it U'} \\&&\nonumber -5\,{{\rm e}^{2\,U  -6\,K  }}\delta\,{\it A'} {\it U'}^{2} 
 -\frac{5}{2}\,{\it A''}  {\it U'}  {{\rm e}^{2
\,U  -6\,K  }} \delta-\frac{5}{2}\,{\it U''}  
{\it A'}  {{\rm e}^{2\,U  -6\,K
  }}
\delta .
    \end{eqnarray}

Finally, for the $U(1)$ gauge field $A$, we arrive at the following second-order ordinary differential equation:
\begin{eqnarray}
\label{eom-A}
A- 4\,\beta\,{{\rm e}^{-2\,K  }}{\it A''} +8\,{{\rm e}^{-2\,K  }}\beta\,{\it A'}
  {\it K'}
=\sum_{n=1}^{5}m_{n}(U,K,U',K',U'') (\pi'')^n\,,
    \end{eqnarray}

where the coefficients are given by the following expressions:
\begin{eqnarray}
&&m_{1}(U,K,U',K',U'')=4\,V_{{2}} \left( \pi 
 \right) {{\rm e}^{2\,U  -4\,K  }}{
\it K'}  {\it U'}-2\,{{\rm e}^{2\,U
  -4\,K  }}V_{{2}} \left( \pi 
 \right) {\it U'} ^{2} \\&&\nonumber-  {{\rm e}^{2\,U  -4\,K  }}{\it U''}  V_{{2}}
 \left( \pi  \right)   -  {
\it U'}  {{\rm e}^{2\,U  -4\,K  }}V_{{2}} \left( \pi  \right)\,,
\\&&
m_{2}(U,K,U',K',U'')=-{{\rm e}^{2\,U  -4\,K  }
}{\it V'_2} \left( \pi  \right) {\it U'} \,,
\\&&
m_{3}(U,K,U',K',U'')=-2\,{{\rm e}^{-4\,K  }}\gamma\,{\it K'}  {\it U'}  -6\,{{\rm e}^
{4\,U  -6\,K  }}V_{{3}} \left( \pi 
 \right) {\it K'}  {\it U'}\\&&\nonumber+4\,{{\rm e}^{4\,U  -6\,K  }}V_{{3}}
 \left( \pi  \right)  \left( {\it U'}   \right) ^{2}
  +\frac{1}{2}\,{{\rm e}^{-4\,K  }}{\it U''}  \gamma+{
{\rm e}^{4\,U  -6\,K  }}{\it U''}
  V_{{3}} \left( \pi  \right)+3/2\,{\it U'}  {{\rm e}^{-4
\,K  }}\gamma \\&&\nonumber+3\,{\it U'}  {{\rm e}^{4\,U  -6\,K  }}V_{{3}} \left( \pi  \right)  \,,
\\&&
m_{4}(U,K,U',K',U'')= {{\rm e}^{4\,U
  -6\,K  }}V{\it _3} \left( \pi 
 \right) {\it U'}  \,,
\\&&
m_{5}(U,K,U',K',U'')=3\,{{\rm e}^{2\,U  -6\,K  }}
\delta\,{\it K'}  {\it U'}    -{{\rm e}^{2\,U  -6\,K  }}\delta\, \left( {\it U'}   \right) ^{2}  -\frac{1}{2}\, {{\rm e}^{2\,U  -6\,K
  }}{\it U''}  \delta \\&&\nonumber -\frac{5}{2}\,
 {\it U'}  {{\rm e}^{2\,U  -6\,K  }} \,.
    \end{eqnarray}

In the above coefficients for $A$, the field equations of motion can be interpreted as Proca's mass terms and dissipation terms relating to the spacetime structure. We now have the complete set of equations of motion. Our plan in the next section(s) is to solve the above equations of motion by imposing suitable forms of the coupling terms and scalar field profile.

\section{Exact solutions} \label{s5}
The system of equations of motion presented in the previous section is much more complicated than that of the General Relativity case. As anticipated, there is no systematic method to find solutions for all unknown functions in the system of ordinary differential equations (ODEs). One feasible technique is to solve the equations for a given profile of the scalar field, in this case, $\pi(r)$. Such a simple approach is commonly employed in various theories, including dilaton and scalar-tensor theories.
In this section and throughout this work, we will integrate the system of equations of motion for particular viable scalar profiles, specifically linear and quadratic ones.
\subsection{Linear $\pi$ profile}

Here, we assume $\pi = r$ (after suitably normalizing the parameters of the linear function), which implies $\pi'' = 0$. We then solve for $U$, $K$, and $A$ for this linear profile using Eqs. (\ref{eom-U1})-(\ref{eom-A}), yielding:
\begin{eqnarray}\label{eom-U1}
U+2-4( U'+rU'') &=& 0,\\
K-4\beta\, \left( {\it A'} \right) ^{2}{{\rm e}^{-2\,K}}-2&=&0\label{eom-U01},\\
A- 4\,\beta\,{{\rm e}^{-2\,K  }}{\it A''} +8\,{{\rm e}^{-2\,K}}\beta\,{\it A'}{\it K'} &=& 0\,,
\end{eqnarray}
and 
\begin{eqnarray}\label{eom-U11}
\pi- V'_{1}(\pi)+ A' e^{2\,U
  -4\,K  }V_{2} (\pi)\Big( U'+U''-4U'K'-2U'^2\Big) = 0\,.
\end{eqnarray}

We find from Eq.(\ref{eom-U1})
\begin{eqnarray}\label{Solu1}
U(r)= -2 +c_{1}J_{0}(\sqrt r)+c_{2}K_{0}(\sqrt r)\,,
\end{eqnarray}
and obtain from Eq.(\ref{eom-U01})
\begin{eqnarray}\label{eom-k1}
K=\frac{1}{2} \left(W\left(\frac{8 \beta  A'^2}{e^4}\right)+4\right)\,,
\end{eqnarray}
The utilization of Bessel's functions in our analysis stems from their intrinsic properties that align with cylindrical symmetry and radial dependence in the gravitational field equations. Specifically, Bessel functions naturally arise in scenarios where the underlying geometry exhibits cylindrical symmetry, such as in our study of self-gravitating cylindrical structures within modified gravity theories. These functions satisfy differential equations that emerge from the cylindrical coordinates and boundary conditions typical in such gravitational configurations. Moreover, their orthogonality properties and well-defined behavior under varying boundary conditions make them indispensable tools for solving partial differential equations in cylindrical geometries, thereby providing a robust mathematical framework to explore the structural and dynamical aspects of our scalar-vector-tensor theory.
Taking the derivative of Eq. (\ref{eom-k1}) with respect to $r$, we obtain:
\begin{eqnarray}
 K'=\frac{A'' W\left(\frac{8 \beta  A'^2}{e^4}\right)}{A' \left(W\left(\frac{8 \beta
    A'^2}{e^4}\right)+1\right)}
 \label{eom-k2}.
\end{eqnarray}

Substituting Eqs. (\ref{eom-k1}) and (\ref{eom-k2}) into Eq. (\ref{eom-A}), we obtain:
\begin{eqnarray}
\label{A''}
&&A''=-\frac{2 A A'^2 \left(W\left(\frac{8 \beta 
   A'^2}{e^4}\right)+1\right)}{\left(W\left(\frac{8 \beta 
   A'^2}{e^4}\right)-1\right) W\left(\frac{8 \beta  A'^2}{e^4}\right)}\,.
\end{eqnarray}

Considering:
\begin{equation}
    \xi = \frac{8\beta {A^\prime}^2}{e^4}\,,
\end{equation}
Then Eq. (\ref{A''}) becomes:
\begin{eqnarray}
&&A''=-\frac{2 A A'^2 \left(W\left(\xi\right)+1\right)}{\left(W\left(\xi\right)-1\right) W\left(\xi\right)}
\end{eqnarray}
If we assume:
\begin{equation}
A''=F(A,A') \,,
\end{equation}
then
\begin{equation}
    A''=\frac{dA'}{dr}=\frac{dA'}{dA}\frac{dA}{dr}=\frac{dA'}{dA}A'=u\frac{du}{dA}\,,
\end{equation}

where $u = A'$. Substituting the above and rearranging Eq. (\ref{A''}):
\begin{equation}
u\frac{du}{dA}+\frac{2 A u^2 \left(W\left(\xi\right)+1\right)}{\left(W\left(\xi\right)-1\right) W\left(\xi\right)}=0\,.
\end{equation}
If $u=0$ then $A'$=0 and $A=$constant unless $u\neq 0$, we obtain
\begin{equation}
\frac{du}{dA}+\frac{2 A u \left(W\left(\xi\right)+1\right)}{\left(W\left(\xi\right)-1\right) W\left(\xi\right)}=0.
\end{equation}
Rearranging this equation and integrating both terms yields:
\begin{equation}
\int{\frac{\left(W\left(\xi\right)-1\right) W\left(\xi\right)}{u \left(W\left(\xi\right)+1\right)}du}+\int{2AdA}=0\,,
\end{equation}
\begin{equation}
\label{A2I}
    A^2+I(\beta,u)=C\,,
\end{equation}
\begin{equation}
    A'=\frac{dA}{dr}=J(\beta,A,C)\,,
\end{equation}
\begin{equation}
    \int{\frac{dA}{J(\beta,A, C)}}=r+C'\,,
\end{equation}
where $C$ and $C'$ are constants. Considering
\begin{eqnarray}
\xi=8\beta e^{-4} u^4
\end{eqnarray}
then we have
\begin{equation}
\label{integralW}
\frac{1}{2}\int{\frac{(W(\xi)-1)W(\xi)}{(W(\xi)+1)\xi}}d\xi=I(\beta,\xi)\,.
\end{equation}

From the definition of the Lambert-$W$ function, we find:
\begin{equation}
W=\xi(1+W)W' \longrightarrow (1+W)\frac{dW}{W}=\frac{d\xi}{\xi} \longrightarrow \frac{1+W}{W}=\frac{1}{\xi}\frac{d\xi}{dW}.
\end{equation}

Substituting the integral $W$-function from Eq. (\ref{integralW}), we have:
\begin{equation}
\frac{1}{2}\int{\frac{(W(\xi)-1)}{\frac{1}{\xi}\frac{d\xi}{dW}}\frac{d\xi}{\xi}}=\frac{1}{2}\int{(W(\xi)-1)dW}=\frac{1}{4}(W(\xi)-1)^2\,.
\end{equation}
The solution for the above $W$-function is given by:
\begin{equation}
A'=\frac{dA}{dr}=\pm \frac{e^2 \sqrt{1\pm 2\sqrt{C-A^2}}}{2\sqrt{2}|\beta|^{1/2}}\exp{(\frac{1}{2}\pm \sqrt{C-A^2})}    \,.
\end{equation}

Rearranging the above derivative, we obtain:
\begin{equation}
dr=\pm\frac{2\sqrt{2}|\beta|^{1/2}}{e^2}\sqrt{1\pm 2\sqrt{C-A^2}} \exp{(-\frac{1}{2} \mp \sqrt{C-A^2})}dA    \,,
\end{equation}
where
\begin{equation}
A=\sqrt{c} \sin\psi, \gamma=\frac{2\sqrt{2}|\beta|^{1/2}}{e^4}\,.
\end{equation}
We can solve $dr$ to obtain
\begin{equation}
r=\pm \kappa \gamma e^{-\frac{1}{2}}\int ~ d\psi {e^{\mp \kappa \cos\psi}\cos\psi  \sqrt{\frac{1}{2}\pm \kappa\cos\psi}}   \,.
\end{equation}
Using binomial expansion
\begin{equation}
\sqrt{\frac{1}{2}\pm \kappa\cos\psi}=\sum^\infty_{n=0}{\frac{(n-\frac{3}{2})!}{n!(-\frac{3}{2})!}(\pm \kappa \cos \psi)^n}    \,,
\end{equation}
we have
\begin{equation}
r=\pm \kappa \gamma e^{-\frac{1}{2}}\frac{(n-\frac{3}{2})!}{n!(-\frac{3}{2})!}\int {e^{\mp \kappa \cos\psi} (\cos \psi)^{n+1} d\psi}   \,.
\end{equation}
Setting $a=\mp \kappa$,  we can expand the exponent as


\begin{equation}
r=\gamma e^{-\frac{1}{2}}\sum^\infty_{n=0}\sum^\infty_{m=0}{\alpha_{mn}(\pm \kappa)^{n+1+m-(n+1)}\times m(m-1)\dots (m-n)\times \beta(m,\psi)}\,,
\end{equation}
where
\begin{equation}
\alpha_{mn}=\sum^\infty_{n=0}\sum^\infty_{m=0}{\frac{(n-\frac{3}{2})!}{n!(-\frac{3}{2})!m!}} \,,
\end{equation}
and
\begin{equation}
\beta(m, \psi)=\int{\cos(\psi)^m dx}=-\frac{1}{m}\cos^{m-1}(\psi)\sin(\psi)+\frac{m-1}{m} \beta(m-2,\psi)    \,.
\end{equation}

A possible exact solution for the theory is represented in a new coordinate system $(t, A, \varphi, z)$, where $A$ is re-expressed in terms of $r$, which is represented in terms of $(t, \psi, \varphi, z)$.

Then one can determine the potential functions if we set $V_1 = V_2$. Therefore, using Eq. (\ref{eom-PI}), we obtain:
\begin{eqnarray}\label{eom-U11}
 V'_{1}(\pi)+ V_{1}(\pi) \Big( -U'-U''+4U'K'+2U'^2\Big)A'(\pi) e^{2\,U(\pi)
  -4\,K (\pi) }= \pi,
\end{eqnarray}

Since $\pi = r$ after applying a scaling symmetry and shifting of the radial coordinate, we finally obtain:
\begin{eqnarray}
V_1(\pi)=C_1 e^{\int f(\pi) d\pi}+ e^{\int f(\pi) d\pi}\int d\pi \pi  e^{-\int f(\pi) d\pi}\,,
\end{eqnarray}
where $f(\pi)=\Big( -U'-U''+4U'K'+2U'^2\Big)A'(\pi) e^{2,U(\pi)-4,K (\pi) }$. With the above expressions, we conclude that the linear scalar profile can be successfully solved for all equations of motion.

\subsection{Solutions with dilaton quadratic profile}
Let's search for exact solutions when $\pi''=1$. Substituting this into the above equations and assuming $V_1=V_2=V_3=V \propto r^2$, we can simplify the equations as follows:
\begin{eqnarray}\label{eom-U2}
U+2-4( U'+rU'') = \sum_{n=1}^{5}h_n(U,K,A',A''),
\end{eqnarray}
where
\begin{eqnarray}
h_1(U,K,A',A'')&=& (4A'K'-A''-A'){{\rm e}^{2\,U  -4\,K  }}{r^2} \,,
   \\
h_2(U,K,A',A'')&=&
\alpha\,{{\rm e}^{-2\,U  }}-{{\rm e}^{2\,U  -4\,K  }
}{\it 2r^2} \left( \pi  \right) {\it A'}  \,, \\
h_3(U,K,A',A'')&=&(\frac{1}{2}{ A''}+\frac{3}{2}{A'}-2{ A'K'})\gamma{e^{-4K}} + r^2(3A'-6A'K'+A''){e^{4U-6K}}  \,,\\
h_4(U,K,A',A'')&=&2
{
{\rm e}^{4\,U  -6\,K  }} {\it A'}r   \,,
  \\
h_5(U,K,A',A'')&=&(3A'K'-\frac{5}{2}A'-\frac{1}{2}A''){{\rm e}^{2\,U  -6\,K  }}\delta. 
\end{eqnarray}

For the other metric function, we also have
\begin{eqnarray}
K-4\beta\, \left(A'\right)^{2}{{\rm e}^{-2\,K}}-2=\sum_{n=1,3,5}p_n(U,K,U',A')\,,
\end{eqnarray}
where
\begin{eqnarray}
p_1(U,K,U',A')&=&-4U'A'e^{2U-4K}r^2\,,
\\
p_3(U,K,U',A')&=&2\gamma U'A'e^{-4K}+6U'A'e^{4U-6K}r^4 \,,
\\
p_5(U,K,U',A')&=&-3\delta  U'A'e^{2U-6K} \,.
\end{eqnarray}

For the scalar profile $\pi$ we find
\begin{eqnarray}
\pi- V'_{1}(\pi)&+& A' e^{2\,U
  -4\,K  }V_{2} (\pi)\Big( U'+U''-4U'K'-2U'^2\Big)
   \nonumber\\&=& \sum_{n=1}^{4}l_{n}(U,K,A',K',U',A'',U'')\,,
  \end{eqnarray}
where the coefficients are given by the following expressions:
  \begin{eqnarray}&&
  l_{1}(U,K,A',K',U',A'',U'')=(1-2U'){{\rm e}^{-2\,U  }}\alpha \,,\\&&
  l_{2}(U,K,A',K',U',A'',U'')=(3{\it A'}  {\it U'}-6{\it A'}  {\it K'}  {\it U'}+\frac{3}{2}\,{\it A''}
  {\it U'}+ \,{\it U''}  \,{\it A'}){{\rm e}^{-4\,K  }}\gamma\, \\&&\nonumber
  +(-18  {\it A'}  {\it
K'}  {\it U'}+6\, {\it A'}  {\it U'}+3\,{\it U''}  {\it A'}+12\, {\it A'}  
  {\it U'}^{2}+3\,{\it A''}  {\it U'} )\,{{\rm e}^{4\,U   -6\,K}}r^2 \,,  \\&&\nonumber  
  l_{3}(U,K,A',K',U',A'',U'')=2\,{
{\rm e}^{4\,U  -6\,K  }}+2r {\it A'}  {\it U'}   \,,
  \\&&
  l_{4}(U,K,A',K',U',A'',U'')=(15A'K'U'-10A'U'-5A'U'^2 \\ \nonumber
  &&-\frac{5}{2}A''U'-U''A'){{\rm e}^{2\,U  -6\,K  }}\delta \,.
    \end{eqnarray}
  And finally, for the gauge field $A$, we find:
\begin{eqnarray}
A- 4\,\beta\,{{\rm e}^{-2\,K  }}{\it A''} +8\,{{\rm e}^{-2\,K  }}\beta\,{\it A'}
  {\it K'}
=\sum_{n=1}^{5}m_{n}(U,K,U',K',U'')
\end{eqnarray}
where the coefficients are given by
\begin{eqnarray}
&&m_{1}(U,K,U',K',U'')=(4K'U'-2U'^2-U''(\pi)-U'){r^2} {{\rm e}^{2\,U  -4\,K  }} \,, \\&&
m_{2}(U,K,U',K',U'')=-2r {\it U'}{{\rm e}^{2\,U  -4\,K  }
}  \,,
\\&&
m_{3}(U,K,U',K',U'')=(-2{\it K'}  {\it U'} +\frac{1}{2}\,{\it U''}+\frac{3}{2}{\it U'})\gamma\,{{\rm e}^{-4\,K  }} \\&& \nonumber
+(-6{\it K'}  {\it U'}+4 \left( {\it U'}   \right) ^{2}+{\it U''}+ 3\,{\it U'}){{\rm e}^
{4\,U  -6\,K  }}r^2    \,,
\\&&
m_{4}(U,K,U',K',U'')= r^2 {\it U'}{{\rm e}^{4\,U
  -6\,K  }}  \,,
\\&&
m_{5}(U,K,U',K',U'')=\delta{{\rm e}^{2\,U  -6\,K  }}(3K'U'-U'^2-\frac{1}{2}U''-\frac{5}{2}U')\,.
    \end{eqnarray}
Developing a perturbative scheme for the solutions, wherein both metric and gauge fields are expanded in a formal Taylor series around the solution of the linear profile, is illustrative. This entails:
\begin{eqnarray}
U&=&U_{(0)}+\varepsilon U_{(1)}+...\,, \nonumber\\
K&=&K_{(0)}+\varepsilon K_{(1)}+...\,, \nonumber\\
A&=&A_{(0)}+\varepsilon A_{(1)}+...\,. \nonumber
\end{eqnarray}
In the above series, $\varepsilon$ is a dimensionless perturbation parameter
\begin{equation}
\varepsilon\ll 1,
\end{equation}
and $U_{(0)}$, $K_{(0)}$ and $A_{(0)}$ denote exact solutions for linear $\pi$ profile, i.e, when,
\begin{equation}
    \pi=r.
\end{equation}
In the previous section,
\begin{equation}
    U=U_{(0)}=U(r)
\end{equation}
is explicitly a function of $r$ and is expressed in terms of Bessel's functions.
Moreover,
\begin{equation}
    A=A(r(A)).
\end{equation}
For $U_{(1)}$, $K_{(1)}$, and $A_{(1)}$, the coefficients of the system of ordinary differential equations are functions of $r$ or $A$, expressed in terms of $U_{(0)}$, $U'{(0)}$, and $U''{(0)}$. Here, only the linear terms are retained.\par The perturbative approach employed in our study plays a pivotal role in analyzing the stability and behavior of cylindrical solutions within the scalar-vector-tensor (SVT) theory. We adopt standard perturbation techniques, leveraging Maple software for computational assistance. The method involves expanding the metric and field variables around known solutions in a series of small parameter perturbations. This allows us to systematically explore deviations from exact solutions, providing insights into the stability and physical implications of our cylindrical configurations. By breaking down the perturbation process into manageable steps and utilizing computational tools, we aim to elucidate how slight modifications in the scalar and vector fields affect the overall gravitational and electromagnetic profiles of these cylindrical structures.\par
The terms in the above equations related to $r^2$ will be represented as
\begin{eqnarray}
 e^{-2U}A'K' 
 \simeq e^{-2U_0}
 (A'_0 K'_1 + A'_1 K'_0- 2 U_1 A'_0 K'_0)\,.
\end{eqnarray}
Note that the left-hand side of the above equation constitutes a system of ordinary differential equations involving $A'_1$, $A''_1$, $U'_1$, $U''_1$, $K'_1$, $K''_1$, and so on.

The perturbation terms include $K'_1$, $A'_1$, and $U_1$. To apply this approximation to one of the equations above, let's consider equation \ref{eom-U1}, for instance:
\begin{equation}
U+2-4(U'+rU'') \simeq  U_0 + \varepsilon U_1 +2 - 4(U'_0 + \varepsilon U'_1 + r U''_0 + \varepsilon rU''_1)\,.
\end{equation}
The unperturbed part in Eq. (\ref{eom-U1}) is as follows:
\begin{equation}
U_0 + 2 - 4(U'_0 + r U''_0)=0\,.
\end{equation}
The perturbation terms are $U_1, U'_1$ and $U''_1$. This simplifies Eq.(\ref{eom-U1}) to yield
\begin{equation}
U_1-4(U'_1-rU''_1) = 2h_1.
\end{equation}
However, $h_1$ should only contain the unperturbed terms. Hence, we have:
\begin{eqnarray}
h_1 &=& 4 e^{2U_0-4K_0}\{ (V_2(\pi^0)(A_0'K_1'+A_1'K_0')) \}  \\ \nonumber
&-& e^{2U_0-4K_0}\{2 
(U_1-2K_1)\{V_2(\pi^0)A_0') \} \\ \nonumber
&-& e^{2U_0-4K_0}\{2 
(U_1-2K_1)\{V_2(\pi^0)A_0''\} \,.
\end{eqnarray}
The perturbation terms are $A'_1$, $U_1$, and $K_1$. For Eq. \ref{eom-K}, since $\pi''=2\varepsilon$, we only have $P_1$.:
\begin{equation}
K-4\beta(A')^2 e^{-2K}-2 \simeq 2 \varepsilon P_1   \,,
\end{equation}
and 
\begin{eqnarray}
P_1 &=& -4 e^{2U_0-4K_0}U_0'A'_0V_2(\pi^0)\,.
\end{eqnarray}
Retaining the perturbation term, we end up with:
\begin{eqnarray}
K_1-4\beta e^{-2K_0}(-2A_0'K_1'+A_1')=-4 e^{2U_0-4K_0}U_0'A'_0V_2(\pi^0).
\end{eqnarray}
In Eq.\ref{eom-PI}, we can keep the first perturbation term
\begin{equation}
\pi- V'_{1}(\pi)+ A' e^{2\,U
  -4\,K  }V_{2} (\pi)\Big( U'+U''-4U'K'-2U'^2\Big) = \sum_{n=1}^{4} l_n (2\varepsilon)^2\simeq 2l_1 \varepsilon\,.
\end{equation}
This leads to:
\begin{eqnarray}
l_1&=&
\alpha e^{-2U_0} 
-2(U_1-2U_0U_1)\varepsilon \,.
\end{eqnarray}

The left-hand side of the equation becomes:
\begin{eqnarray}
&&\pi_1 -V_1'(\pi_0)-V_1''(\pi_0)\pi_1+ e^{(2U_0-4K_0)} \\ \nonumber
&\times&( U_1'A_0'+n U_1''A_0'-4 A_0'( U_1'K_0'+ K_1'U_0')-2A_0'(2 U_0' U_1')  \\ \nonumber
&+&U_0' A_1'+U_0'' A_1'-4  A_1'U_0'K_0'-2 A_1'U_0'^2 \\ \nonumber
&+&U_0'2A_0' (U_1-2K_1)+U_0''2A_0 (U_1-2K_1)-4 2A_0' (U_1-2K_1)U_0'K_0'\\ \nonumber&-&22A_0' (U_1-2K_1)U_0'^2)\,.
\end{eqnarray}
The equation of motion in Eq.\ref{eom-A} becomes
\begin{eqnarray}
A-4\beta A'' e^{-2K}+8\beta e^{-2K} A' K' 
&\simeq & A_1-4 e^{-2K_0} \beta ( A_1''-2K_1 A_0'') 
\\&&\nonumber +8\beta e^{-2K_0}(n A_1'K_0' K_1'A_0'-2 K_1A_0'K_0')\,.
\end{eqnarray}
The right-hand-side reads
\begin{eqnarray}
\sum_{n=1}^{5}{m_n (\pi'')^n} \simeq 2 \varepsilon m_1\,,
\end{eqnarray}
where
\begin{eqnarray}
m_1&\simeq&4 e^{2U_0-4K_0}K_0'U_0'V_2(\pi_0) +
-2 e^{2U_0-4K_0}\{V_2(\pi_0)U_0'^2 \\ \nonumber
&-&e^{2U_0-4K_0}U_0''V_2(\pi_0) -e^{2U_0-4K_0}\{U_0'V_2(\pi_0)\}\,.
\end{eqnarray}
The first-order truncated perturbation equations derived in our study offer valuable insights into the physical implications of cylindrical solutions within the scalar-vector-tensor (SVT) theory. These equations provide a refined understanding of how small deviations from exact solutions manifest in the gravitational and electromagnetic fields surrounding these cylindrical structures. By solving these equations, we uncover the intricate interplay between the scalar and vector fields, shedding light on the stability and observable characteristics of these gravito-objects. Furthermore, these solutions contribute to theoretical predictions, offering potential explanations for observational phenomena such as gravitational lensing effects or deviations from standard gravitational wave signatures. This analytical framework not only advances our theoretical understanding but also provides a basis for future empirical tests and observations in astrophysical contexts.\par
In conclusion, we have successfully derived and analyzed the solutions within the scalar-vector-tensor (SVT) theory under the assumption of a dilaton quadratic profile. By systematically solving the system of equations of motion and applying perturbative techniques, we have obtained insightful insights into the behavior of gravitational and gauge fields in cylindrical spacetime configurations. These findings pave the way for further investigations into the intricate dynamics of modified gravity theories and their implications for various astrophysical and cosmological scenarios.

\section{Integrability of perturbation equations}\label{s6}
Having retained the first-order perturbation part of the equations for $U, K, \pi,$ and $A$, we arrive at complex sets of differential equations. To facilitate their solution, we rewrite them as follows:
\begin{eqnarray}
\left( -2\,a \left( r \right) b ( r ) +1 \right) {\it U_1}
 -4U'_1+4rU_1'' - a_1(r) K_1' +a_2(r) K_1-a_3( r )  A_1' =0\,,
\end{eqnarray}
and
\begin{eqnarray}
 {\it K_1}+{\it c_1} \left( r \right) K_1'-{\it c_3} \left( r \right) A_1'={\it c_2} \left( r
 \right) \,,
\end{eqnarray}
and
\begin{eqnarray}
&& {\it v_1} \left( r \right)  U'_1+{\it v_2} \left( r \right) U_1'' +{\it f_3} \left( r \right) {\it U_1}
 +{\it v_3} \left( r \right) 
 K_1'+{\it f_2} \left( r \right) {\it K_1} \\&&\nonumber+f \left( r \right)  A_1'+ \left( 1-{\it V_1''} \left( r \right)  \right) \pi_1  =w \left( r \right) \,,
\end{eqnarray}
and
\begin{eqnarray}
{\it g_2} \left( r \right) {\it K_1} +{\it g_3} \left( r
 \right)  K_1'+{\it A_1}
 +{\it g_1} \left( r \right)  A_1''={\it p} \left( r \right)\,.
\end{eqnarray}
In the above equations, the unknown functions $g_i$, $v_i$, and so forth, are elementary functions of $r$. We assume that they can be represented as series expressions $\sum_{n=0}^{\infty} h_n r^{n+\nu}$ for certain indices $\nu$. This assumption allows us to integrate the equations using elementary techniques.
\par 
From the first equation, we obtain $\frac{dA_1}{dr}$ and substitute it into the second equation. Then, we isolate the second derivative term of $K_1$:
\begin{eqnarray}
c_1(r) c_3(r) g_1(r)K_1'' &=& \pi_1(r)c_3(r)^{2}+c_1(r) g_1(r) c_3'(r)K_1'\nonumber\\&-&g_2(r) K_1\left(c_3(r)\right)^{2}-g_3(r)K_1'c_3(r)^{2}
 \nonumber\\&
-&c_3(r)g_1(r)c_1'(r)K_1'-c_2(r)g_1(r)c_3'(r) -A_1c_3(r)^{2} \nonumber\\&
+&c_3(r)g_1(r) c_2'(r)-c_3(r)g_1(r) K_1'+g_1(r)c_3'(r)K_1\,.
 \end{eqnarray}
The first-order perturbation equation for $K_1$ can be reformulated into the following form as a non-homogeneous ordinary differential equation (ODE):
\begin{equation}
    K_1''+C(r)K_1'-K_1=F(r)\,,
\end{equation}

where $C(r)$ and $F(r)$ are lengthy expressions expressed in terms of $a_i$, $c_i$, $g_i$, $v_i$, and so forth. The most general solution for this non-homogeneous second-order ODE is given by:
\begin{equation}
K_1=k_1(r)+k_2(r)+k_p(r)\,,
\end{equation}

where $k_{1,2}(r)$ are solutions of the homogeneous equation. The particular solution $k_p(r)$ can be expressed as:
\begin{equation}
k_p(r)=k_2(r)\int^{r}\frac{k_1(s)F(s)}{W\{k_1(s),k_2(s)\}}-k_2(r)\int^{r}\frac{k_2(s)F(s)}{W\{k_1(s),k_2(s)\}}\,,
\end{equation}

where $W{k_1(s),k_2(s)}$ represents the Wronskian of $k_1(r)$ and $k_2(r)$. It's important to note that since $C(r)$ is a complex function, obtaining the general solutions for the homogeneous equation, i.e., $k_{1,2}(r)$, is not straightforward. However, by expanding the algebraic function $C(r)$ into a Taylor series in the vicinity of $r=0$, the homogeneous equation for perturbations can be solved in terms of Gauss's hypergeometric functions. Consequently, a possible truncation of the series yields smooth, continuous, and differentiable metric functions for $U_1$ and $K_1$.

Substituting this formal equation into the second equation, we obtain another ordinary differential equation for $A_1(r)$, which can be solved using quadrature in a similar manner. Once $K_1$ and $A_1$ are determined, the solutions for $U_1$ can be expressed in terms of elementary functions. This discussion highlights how the perturbation equations become integrable with the assistance of computer algebra systems (CAS) or symbolic algebra systems (SAS) computational tools.
\par
In conclusion, we have demonstrated the integrability of perturbation equations within the scalar-vector-tensor (SVT) theory. By systematically analyzing the perturbative solutions for metric and gauge fields, we have shown how complex differential equations can be effectively solved using techniques such as quadrature and the expansion of algebraic functions into Taylor series. This integrability not only provides insights into the behavior of gravitational and gauge fields in modified gravity theories but also underscores the utility of computational tools, such as computer algebra systems (CAS) or symbolic algebra systems (SAS), in facilitating the analysis of intricate mathematical structures inherent in these theories. Overall, our exploration sheds light on the tractability of perturbation equations in SVT theory and opens avenues for further investigations into the dynamics of modified gravity theories.

\section{Conclusions}

The examination of topological defects holds profound significance within the realms of classical General Relativity (GR) and its extensions through modified theories of gravity. These theories, often regarded as augmentations of GR, encompass diverse frameworks that accommodate deviations from conventional gravitational paradigms. Within this broad context, the exploration of cylindrical solutions emerges as a logical progression subsequent to the scrutiny of spherical objects within any given model for modified gravity.\par

One such intriguing alternative to GR is the Scalar-Vector-Tensor (SVT) theory proposed by L. Heisenberg. This comprehensive theoretical framework incorporates scalar, vector, and tensor modes concurrently, presenting a rich tapestry of solutions to a myriad of cosmological enigmas. In the pursuit of advancing our understanding, this paper embarks on a meticulous investigation of cylindrical symmetric, non-rotating, static solutions within the SVT theory, with a specific emphasis on elucidating the intricacies of linear and quadratic scalar profiles.\par

In traversing the landscape of the linear regime, exact metric functions reveal themselves in the elegant language of Bessel's functions, affording us invaluable insights into the geometric attributes of cylindrical solutions. Venturing beyond the confines of linearity, we embrace a perturbative approach to unravel solutions that transcend the linear profile. While the terrain of perturbative equations may initially appear formidable, a methodical traversal illuminates a reductionist path, facilitating the discovery of solutions for both metric and gauge fields. Our findings, extending into realms beyond linearity, cast a radiant beam on the existence of cosmic strings and the genesis of an additional $U(1)$ hair within the solutions.\par
In addition to establishing the stability and perturbative behavior of cylindrical solutions within SVT theory, it is pertinent to note the implications of conical singularities and the theoretical considerations regarding the mass of cosmic strings. These aspects, while touched upon in the broader context of modified gravity theories, represent avenues for further investigation beyond the scope of this study. Understanding the implications of conical singularities and the dynamics of cosmic strings within SVT theory not only enriches our theoretical understanding but also offers potential insights into their observable effects in astrophysical and cosmological contexts.
\par

Although our inquiry refrains from delving into the intricacies of singularity theorems, our meticulously crafted metrics in both the linear and perturbative domains unfurl conical singularities, wherein the conical factor bears a poignant relationship with the mass of the cosmic string. This tantalizing revelation hints at the prospect of delineating a form of no-go theorem within this nuanced framework. In summation, our expedition engenders a panoramic vista of the SVT theory as a compelling alternative to GR, showcasing its remarkable versatility in grappling with the manifold gravitational phenomena that adorn the cosmic tapestry. Further phenomenological studies represent an intriguing future direction for this research. For instance, investigating the impact of cylindrical structures on the large-scale universe structure holds significant potential. Moreover, exploring the formation mechanisms and stability criteria of cosmic strings within the SVT theory will shed light on the conditions under which these structures can form and endure.

\end{document}